\documentclass[10pt, conference, compsocconf]{IEEEtran}
\usepackage[T1]{fontenc}
\usepackage[figure,linesnumbered]{algorithm2e}
\usepackage{amsmath}
\usepackage{booktabs}
\usepackage{cite}
\usepackage{color}
\usepackage{comment}
\usepackage{enumerate}
\usepackage{epsfig}
\usepackage{float}
\usepackage{graphicx}
\usepackage{latexsym}
\usepackage{mathptmx}
\usepackage{multirow}
\usepackage{soul}
\usepackage{tabularx}
\usepackage{times}
\usepackage[hyphens]{url}
\usepackage{wrapfig}
\usepackage{listings}
\usepackage{lipsum}
\usepackage[labelformat=simple]{subcaption}
\usepackage{algorithmic}
\usepackage{textcomp}
\usepackage{blindtext}
\usepackage{caption}
\usepackage{subcaption}

\title{Multi-Path Routing on the Jellyfish Networks}

\author{
\IEEEauthorblockN{
	Zaid ALzaid, Xin Yuan, Saptarshi Bhowmik}
\IEEEauthorblockA{
	Department of Computer Science\\
	Florida State University\\
	Tallahassee, Florida, USA\\
\{alzaid, xyuan, mhowmik\}@cs.fsu.edu}
}

\begin{document}

\maketitle

\begin{abstract}
  The Jellyfish network has recently be proposed as an alternate to the fat-tree
  network as the interconnect for data centers and high performance computing clusters.
  Jellyfish adopts a random regular graph as its topology and has been showed to be more
  cost-effective than fat-trees. Effective routing on Jellyfish is challenging. It is
  known that shortest path routing and equal-cost multi-path routing (ECMP) do not work
  well on Jellyfish. Existing schemes use variations of k-shortest path routing (KSP).
  In this work, we study two routing components for Jellyfish: path
  selection that decides the paths to route traffic, and routing mechanisms that
  decide which path to be used for
  each packet. We show that the performance of the existing KSP
  can be significantly improved by incorporating two heuristics, randomization and
  edge-disjointness. We evaluate a range of routing mechanisms including traffic
  oblivious and traffic adaptive schemes and identify an adaptive routing scheme
  that has significantly higher performance than others including the
  Universal Globally Adaptive Load-balance (UGAL) routing.  
\end{abstract}

\section{Introduction}
\label{intro}


The Jellyfish interconnect has recently been proposed for
data centers and high performance computing (HPC) clusters
\cite{singla2012jellyfish,yuan2013new}.
It adopts the random regular graph (RRG) as its switch-level topology.
Jellyfish has good topological properties such as high bisection bandwidth and low
average path length, and has been shown to be more cost-effective than fat-trees
\cite{singla2012jellyfish}.

Due to the random connectivity in Jellyfish, traditional shortest
path routing and equal-cost multi-path routing (ECMP) 
do not perform well \cite{singla2012jellyfish}. 
Singla et al. proposed k-shortest path routing (KSP) to select paths and
explore path diversity in the network. A variation of KSP, called LLSKR,
was later developed \cite{yuan2013new}. Both KSP and LLSKR suffer from two
issues that can degrade their performance. First, both routing schemes
try to use ``shortest'' paths without considering the potential link usage.
When a link is in a short path,
it tends to be in other short paths as well. Hence, routing traffic with such paths
can introduce load imbalance in the network and lower the performance. Second,
in the jellyfish topology, there are many paths of the same length between each
pair of nodes. The vanilla KSP that is deterministic tends to have biases in
selecting paths, which also results in load imbalance.

Beside {\em path selection} like KSP that decides the set of paths to be used
to route traffic, another important component in routing is deciding
a particular path for each packet. We will call this
component the {\em routing mechanism}. While some conventional routing mechanisms can
apply to multi-path routing on Jellyfish, this compoent has not been
systematically investigated for Jellyfish, and it is unclear whether the
conventional routing mechanisms are effective.

In this work, we study both path selection and routing mechanisms for multi-path
routing on Jellyfish. For path selection,
we find that two heuristics, randomization and edge-disjointness,
can significantly improve the quality of the paths selected when 
they are incorporated in the vanilla KSP. We investigate
a range of routing mechanisms including traffic oblivious and traffic adaptive schemes.
We perform extensive evaluation to compare the path selection
methods and the routing mechanisms using performance modeling, flit-level simulation,
and applications. The main conclusions of this study include the following:

\begin{itemize}
\item For path selection, we find that the paths computed using the KSP
  with randomization and edge-disjointness
  heuristics achieve higher performance than the vanilla KSP across all routing
  mechanisms and all traffic patterns in the study. The improvement is
  significant, especially when the best performing adaptive routing scheme is used. 
  
\item For routing mechanism, we find a new adaptive routing scheme, which we
  call {\em KSP-adaptive} routing, is most effective for multi-path routing on
  Jellyfish. {\em KSP-adaptive} selects a path for each packet by randomly
  obtaining two candidate paths from the $k$ paths for the source-destination pair,
  and choosing the one with a smaller (estimated) latency to route the packet. Our
  evaluation indicates that {\em KSP-adaptive} achieves higher
  performance than various forms of universal globally adaptive load balance
  routing (UGAL) \cite{singh2005load} on Jellyfish.
\end{itemize}

The rest of the paper is structured as follows. Section~\ref{backg}
presents the background. Section~\ref{ours}
describes our new proposed routing schemes.
Section~\ref{perf} reports the performance study.  The related work is discussed in
Section~\ref{related}. Finally, Section~\ref{conc} 
concludes the paper.

\section{Jellyfish}
\label{backg}

Singla et al. \cite{singla2012jellyfish} proposed a jellyfish topology as a
flexible and cost-efficient topology for large scale interconnection networks.
The switch-level topology of Jellyfish is a random
regular graph (RRG), where all switches have the same degree, but are randomly connected.
We consider the case when all switches connected
with the same number of processing nodes and have the same number of ports.

\begin{figure}[htbp]
  \centering
  \includegraphics[width=2.8in]{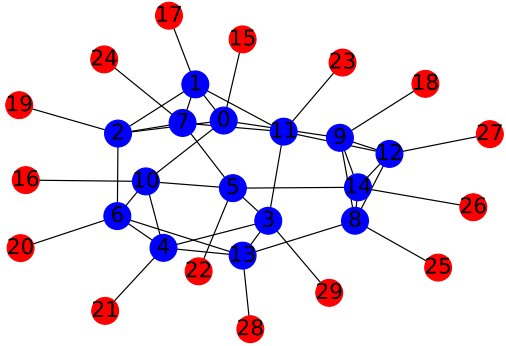}
  \caption{An example $RRG(N=15, x=4, y=3)$}
  \label{fig:JF15}
\end{figure}

Jellyfish can be specified with three parameters \cite{singla2012jellyfish}: the number of
switches ($N$), the number of ports in each switch ($x$), and the number of
ports in each switch that connect to other switches ($y$):
each switch connects to $x - y$ processing nodes. We will use $RRG(N, x, y)$ to denote
a Jellyfish topology with $N$ switches, $x$ switch ports, and $y$ ports in each switch
connecting to other switches. Figure ~\ref{fig:JF15}
illstrates $RRG(N=15, x=4, y=3)$. Each switch
connects to $4-3=1$ processing node. Note that each instance of an $RRG$ is different
from other instance $RRG$. However, when $N$ and $y$ are sufficiently large,
different instance will have very similar network characteristics \cite{yuan2013new}.

Even though RRG provides high network capacity, it is known that 
single path routing and equal-cost multipath routing performs poorly on Jellyfish
\cite{singla2012jellyfish}. As a remedy, Singla et al. \cite{singla2012jellyfish}
proposed the K-shortest path routing (KSP) scheme to explore the path diversity
in the topology. Due to the random nature
of Jellyfish, the lengths of the $K$ paths between different pairs of switches 
may vary. 
Yuan et al. \cite{yuan2013new} observe that using the plain KSP in Jellyfish
has some limitations including (1) not using all ``short'' paths when the number
of short paths between two switches is large, and (2) using long paths when
the number of short paths between two switches is small. Based on the observations,
they proposed Limited Length Spread k-shortest Path Routing (LLSKR), which attempts to
overcome the aforementioned limitations. LLSKR allows a variable number of paths between
two switches, which allows more short paths to be used in comparison to KSP
and can control the number of long paths to be used. 

Both KSP and LLSKR rely on a k-shortest routing algorithm to find the paths.
Finding loopless $k$ shortest paths in a given topology is a generalization of the shortest
path problem. This problem has been studied extensively
\cite{hoffman1959method,yen1971finding,martins2003new}. The baseline algorithm used in this work 
is the Yen's algorithm \cite{yen1971finding}, which is shown
in Figure~\ref{alg:the_alg}.
Since our heuristics are added over this algorithm, we
describe the algorithm for completeness.

\begin{algorithm}[htbp]
  \small
\SetAlgoLined\SetArgSty{}
\KwIn{$Graph$, $Source$ and $Destination$}
\KwOut{K-shortest paths between $Source$ and $Destination$}
\SetKwProg{Fn}{Function}{}{end}
\SetKwIF{If}{ElseIf}{Else}{if}{then}{else if}{else}{endif}
\Fn{searching K-Shortest paths}{
   A = []  a set to store the K- shortest path \\
   B = []  a set to store the potential kth shortest path\\
   
   A[0] = the shortest path from the $Source$ to the $Destination$.\\
  \For{ $i=1$ to $i= k - 1$ }
  {
        \ForEach{ node $j$ in $A[i - 1]$ without the $Destination$ }{
        $ spurNode $ = $j$\\
        $ rootPath $ = $A[i - 1]$ from $Source$ to $ spurNode $ \\

       \ForEach {path $P$ in A}
       {    
           \If{ $rootPath$ = $P[Source:spurNode]$}
           {
           remove edge $P[spurNode:spurNode + 1]$ from the Graph.\\
           }
       }
       \ForEach { $node$ $\in$ $rootPath$ except $spurNode$}
       {
            remove $node$ from the Graph. \\
       }
       $spurPath$ = Dijkstra(Graph, $spurNode$, $Destination$)\\
       $totalPath$ = rootPath +  $spurPath$\\
       \If{ $totalPath \notin B$ }
       {
         add $totalPath$ to $B$
       }

       restore the original Graph.\\
	}
	   \eIf{ $B$ is not empty}
       { 
        $A[K] =$ Shortest path in $B$\\
        $B = []$\\
       }
       {
        end the loop.\\
       }
  }
}
\caption{Yen's algorithm \cite{yen1971finding}}
\label{alg:the_alg}
\end{algorithm}

Yen's algorithm uses two containers $A$ and $B$. Container $A$ stores the $k$-shortest paths
found and container $B$ stores potential shortest paths. The algorithm first finds the shortest
path between the source and destination using Dijkstra's shortest path algorithm (or
any other algorithm that finds a shortest path). The algorithm then loops for $k-1$ iterations,
finding one shortest path in one iteration (Lines 5 to 30). In each iteration,
the algorithm then iterates over all nodes
on the last shortest path, except the destination node. For each node, the algorithm finds the
shortest path to the destination from that node ($spurPath$ in Figure~\ref{alg:the_alg}).
To avoid loop in $spurPath$, the algorithm removes all nodes from the source to the
node before the current node on the shortest path, and the edge from the current node
and the node after it on the shortest path to force the shortest-path algorithm find a
different path as $spurPath$. The algorithms then forms the $totalPath$ by concatenating
the $rootPath$ (the shortest path from the source node to the node before
the current node) and the $spurPath$ as a candidate shortest path. After visiting all
intermediate node along the shortest path, the algorithm selects the shortest path in B
as the next shortest path and move it to A.  The process is then repeated until all
$k$ paths are found.


Besides path selection, the {\em routing mechanism}, which decides the path
for each packet, is also an important compoents in multi-path routing
on Jellyfish. This component has not been sufficiently investigated.
Existing work has assumed some mechanisms that can take advantage of the multiple
paths such as multi-path TCP (MPTCP) \cite{Han06}. There are, however,
other adaptive routing mechanisms, such as the Universal Globally Adaptive
Load-balancing routing (UGAL) \cite{singh2005load}, that has not been carefully studied
in this context. 
  

\section{Multi-path Routing schemes for Jellyfish}
\label{ours}

\subsection{Path selection}

\begin{figure}[htbp]
    \centering
    \begin{subfigure}[t]{0.4\textwidth}
        \centering
        \includegraphics[width=1.5in,height=1.8in]{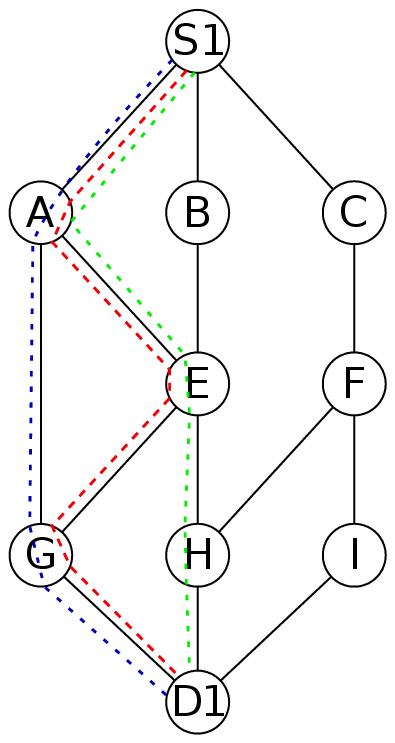}
        \caption{3-shortest paths computed by vanilla KSP}
        \label{fig:SWOL}
    \end{subfigure}
    ~ 
	\begin{subfigure}[t]{0.4\textwidth}
        \centering
        \includegraphics[width=1.5in,height=1.8in]{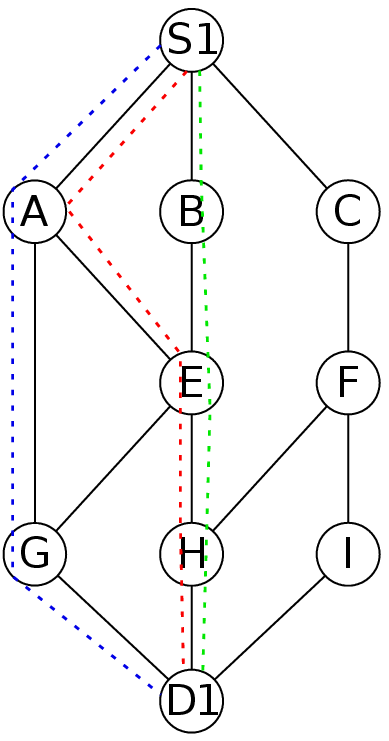}
        \caption{3-shortest paths computed by KSP with randomization}
        \label{fig:PKSP}
    \end{subfigure}
        ~ 
    \begin{subfigure}[t]{0.4\textwidth}
        \centering
        \includegraphics[width=1.5in,height=1.8in]{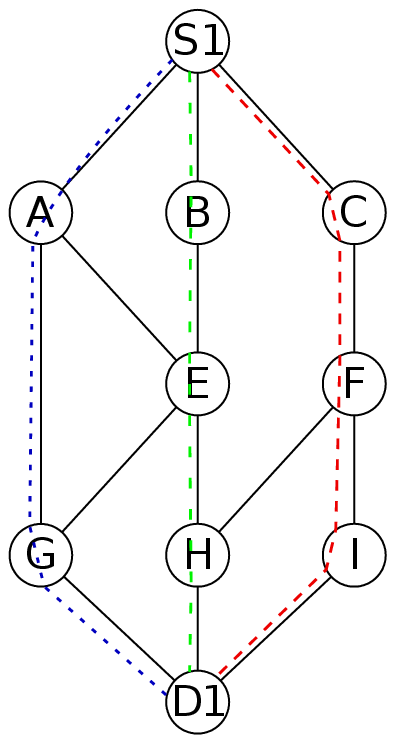}
        \caption{3-shortest paths computed by KSP with edge-disjointness}
        \label{fig:PEDKSP}
    \end{subfigure}
    \caption{Paths computed by KSP with different heuristics}
    \label{fig:SS}
\end{figure}

Both KSP and LLSKR select ``shortest'' $k$-paths. Additionally, the vanilla
version of KSP relies on the Dijkstra's algorithm, and is deterministic: when there are more than one
path that can be selected, the tie is broken deterministically using some means such as 
preferring nodes whose ranks are higher (or lower). The combination can result in low-quality paths
for $k$-path routing in Jellyfish as illustrated in the example in Figure~\ref{fig:SS}.
In this example, we consider KSP that finds 3 shortest paths
from $S1$ to $D1$. From $S1$ to $D1$, there are one 3-hop path
$P0 = S1\rightarrow A\rightarrow G\rightarrow D1$, and
six 4-hop paths,
$P1 = S1\rightarrow A\rightarrow E \rightarrow G\rightarrow D1$,
$P2 = S1\rightarrow A\rightarrow E \rightarrow H\rightarrow D1$,
$P3 = S1\rightarrow B\rightarrow E \rightarrow G\rightarrow D1$,
$P4 = S1\rightarrow B\rightarrow E \rightarrow H\rightarrow D1$,
$P5 = S1\rightarrow C\rightarrow F \rightarrow H\rightarrow D1$, and 
$P6 = S1\rightarrow C\rightarrow F \rightarrow I\rightarrow D1$.

Let us assume that the textbook Dijkstra's algorithm is used in KSP where nodes with smaller
ranks are explored first to find the shortest path. We will call this version
of KSP, the vanilla KSP. The vanilla KSP will first find the three hop path ($P0$).
After that, it will find a 4-hop path with a bias toward nodes with smaller
ranks (in Figure~\ref{fig:SS},
such as a path is the leftmost feasible path).
As a result, the second path that will be found by KSP is
$P1 = S1\rightarrow A\rightarrow E \rightarrow G \rightarrow D1$; and
the third path to be found is $P2 = S1\rightarrow A\rightarrow E \rightarrow
H\rightarrow D1$. These three paths are showed in Figure~\ref{fig:SWOL}.
As can be seen in the figure, all three paths share the link $S1\rightarrow A$. As such,
although there exist three paths to carry traffic from $S1$ to $D1$, the effective bandwidth
from $S1$ to $D1$ is equivalent to only having a single path. Notice 
there typically exist many paths of the same length from a source switch to a destination
switch in an RRG \cite{yuan2013new}. The biases in the deterministic vanilla KSP algorithm
can have serious problems for the Jellyfish topology.

This problem can be alleviated by incorporating randomness in KSP. The randomized KSP
can be achieved by using a randomized Dijkstra's shortest path algorithm:
when exploring the next node, the randomized Dijkstra does not use a deterministic mechanism,
such as the node rank, to break the tie. Rather, it randomly selects a next node when
there are multiple options. Using this heuristic, in the example in Figure~\ref{fig:SS},
after the KSP finds the 3-hop
path, it will randomly select two 4-hop paths out of the 6 candidate paths. For example, it may
select: $P2 = S1\rightarrow A\rightarrow E \rightarrow H\rightarrow D1$ and
$P4 = S1\rightarrow B\rightarrow E \rightarrow H \rightarrow D1$.
This is shown in Figure~\ref{fig:PKSP}. As can be seen from the figure, each link is at most shared
by 2 paths, and the effective bandwidth from $S1$ to $D1$ is equivalent to having 2 independent
paths, which is better than the paths computed by the vanilla KSP.

Sharing links among the selected paths effectively reduces the total achievable
bandwidth from the source to the destination. Randomization does not eliminate link sharing
among the selected paths. The edge-disjoint heuristic completelly eliminates link sharing.
In our implementation, the edge-disjoint heuristic follows the Remove-Find (RF)
method \cite{guo2003link}. The RF method consists of two steps: (1) finding the shortest
path between from the source to the desitnation, and (2) removing all edges associated
with the shortest path. The algorithm repeats these two steps $k$ times or until the source
and destination node are disconnected. Using this heuristic, the 3 paths found for the
example in Figure~\ref{fig:SS} are $P0$, $P4$, and $P6$, as shown in Figure~\ref{fig:PEDKSP}.
As can be seen from this figure, all of the three paths are now link disjointed.
Using the three paths, the total bandwidth from $S1$ to $D1$ is equivalent
to using three independent paths.

There are two potential problems with the edge-disjoint heuristic.
First, there may not exist a sufficient number of edge-disjoint paths between two nodes. Second, 
the lengths of edge-disjoint paths can be much larger than the paths computed by the vanilla KSP,
which increases the network resources to carry a packet and can decrease the performance when
the network is under high load. However, as will be shown in our evaluate, both are insignificant 
for typical Jellyfish networks. The average path length computed
by KSP with the edge-disjoint heuristic is similar to those computed by the vanilla KSP. 

In the rest of the paper, we will use KSP to denote the vanilla KSP, rKSP for the randomized KSP,
EDKSP for the edge-disjoint KSP, and rEDKSP for KSP with both randomization and edge-disjoint
heuristics. Note that rEDKSP will not improve the total throughput for each source-destination pair.
However, randomizing path selection when there are more possibilities results in a better
load balancing for traffic patterns where multiple sources are communicating with
multiple destinations.

\subsection{Routing mechanism}
\label{sec:mechanism}

Given the multiple paths that can be used to carry a packet, the routing mechanism decides
the path for the packet. We investigate a range of routing mechanisms for the multi-path
routing in Jellyfish. In particular, we consider schemes to incoporate the Universal Globally
Adaptive Load-balanced routing (UGAL) \cite{singh2005load}, which has demonstrated effectiveness
on  other topologies such as mesh and Dragonfly
\cite{singh2005load,Kim:2008:ISCA:Dragonfly,Jiang:2009:ISCA:Indirect_adaptive,Faizian:2018:TMSCS,Mollah:2019:TOPC,Rahman:2019:TUR:3295500.3356208}.

UGAL distinguishes between two types of paths, minimal paths and non-minimal paths.
The minimal path is the shortest path from the source to the destination in Jellyfish.
A non-minimal path is formed by two minimal paths, one from the source to a (randomly selected)
intermediate node and the other one from the intermediate node to the destination. 
To route a packet from a source
to a destination, UGAL considers two paths, one minimal path and one random non-minimal path (with
a random intermediate node), compares the estimated packet latency of the two paths
using queue length, and chooses the path with the smaller latency for the packet. 
This scheme, which will be called {\em vanilla UGAL} can be directly applied to Jellyfish.

The vanilla UGAL uses more paths than the paths found by the KSP algorithm. One can restrict
the non-minimal paths to only the one computed by the KSP algorithm. In this case, the shortest
path from the source to the destination will be the minimal path, and the rest of the $k$ shortest
paths are candidates for the random non-minimal paths. We will call this algorithm {\em KSP-UGAL}.
Note that depending on the traffic and network condition, restricting non-minimal paths to
the $k$ shortest path can have advantages over the vanilla UGAL that may use longer non-minimal
paths. In the situation when $k$ paths can provide sufficient path diversity, using short
paths reduces the link usage, which improves the performance.

In the Jellyfish topology, most of $k$ shortest paths are of similar length. Thus,
when only the $k$ paths are used, minimal and non-minimal paths are similar and do not need to be
differentiated. We propose a new adaptive routing scheme, which we call {\em KSP-adaptive}. 
KSP-adaptive randomly selects two paths among the $k$ shortest paths and 
chooses the one with a smaller estimated latency for the packet.

We compare performance of these traffic adaptive routing schemes among one another and against
other traffic oblivious schemes including random routing ({\em random} that randomly selects a path
out of the $k$ paths to route each packet, and {\em round-robin} that uses the $k$ paths for each
source-destination pair in a round-robin fashion to route packets.

\section{Evaluation}
\label{perf}

We perform comprehensive modeling and simulation studies to evaluate the proposed path selection
and routing mechanisms on a number of Jellyfish topologies. In the following, 
we first describe our experimental methodology and then report the evaluation
results. 

\subsection{Methodology}

\subsubsection*{Topology}
We report results on three Jellyfish topologies, a small topology
$RRG(36, 24, 18)$ with 36 switches and 216 compute nodes, a medium sized topology
$RRG(720, 24, 19)$ with 720 switches and 3600 compute nodes,
and a large sized topology $RRG(2880, 48, 38)$ with 2880 swithes and
28800 compute nodes. Table~\ref{tab:JFtopo} summarizes important
parameters of the topologies. 

\begin{table}[htbp]
  \centering
  \begin{tabular}{|l|r|r|r|r|r|}
  \hline
  topology &  switch & No. of & No. of  & average\\
  &  size   & switches & compute  & shortest\\
  &         &          &   nodes & path len.\\
     \hline
     RRG(36,24,16)  &  24       & 36       & 288 & 1.54 \\
     \hline
     RRG(720,24,19)   & 24      & 720      & 3600 & 2.57 \\
     \hline
      RRG(2880,48,38)  & 48     & 2880     & 28800 & 2.59 \\
    \hline
  \end{tabular}
  \caption{Jellyfish topologies used in the experiments}
  \label{tab:JFtopo}
\end{table}

\subsubsection*{Path selection and routing mechanisms}
\label{perf:MPRS}

We compare four different path selection schemes for computing multiple paths
for each pair of source and destination: KSP, rKSP (randomized KSP),
EDKSP (Edge-Disjoint KSP), and rEDKSP (randomized Edge-Disjoint KSP). We will use the
notations KSP(k), rKSP(k), EDKSP(k), and rEDKSP(k) to denote KSP, rKSP, EDKSP, and
rEDKSP with $k$ paths, respectively.
The routing mechanisms considered are the following, which are described in
Section~\ref{sec:mechanism}: single path routing ($SP$),
{\em random}, {\em round-robin}, {\em vanilla UGAL}, {\em KSP-UGAL},
and {\em KSP-adaptive}.


\subsubsection*{Performance Metrics}
We evaluation the path computation and multi-path routing schemes in three different
ways. First, we use a throughput model \cite{yuan2013new} that approximates the
throughput for a given traffic pattern and a topology. Second, 
we use Booksim 2.0 \cite{Jiang:2013:booksim},
  a cycle-accurate interconnection network simulator, to
evaluate the aggregate throughput and packet latency for different schemes.
Finally, we use the CODES 1.0.0 \cite{Cope:2011:Codes}
to measure the communication times for
common communication patterns in HPC applications.

\subsubsection*{Throughput Model}

The throughput model \cite{yuan2013new} estimates the throughput for multipath
routing schemes with multipath TCP (MPTCP \cite{Han06}) like protocol
where each flow realizes by multiple (TCP) sub-flows. 
Given a communication pattern that consists of a set of SD pairs with multipath
routing schemes. The model first computes the number of times each link used by
all sub-flows in the pattern. Assume that for any SD pair in the pattern, there
are $K$ multipath exists. Then the model finds the links load, where for any
link $l$ with capacity $C$ used $X$ times; the link load defined as
$load_{l}$ = $ \frac{X}{C} $. Then the model sets the rate for each sub-flow
in the pattern to be equal to the load of the link with the maximum load along
the path. Finally, the model computes the flow rate by adding up the K sub-flows
rates associated with the flow. 

\begin{equation}
 T(s,d)\sum_{n=1}^{K} \frac{1}{max_{l \in path_{n}(s,d)}}.
\end{equation}

\subsubsection*{Simulator modification and settings}

Both Booksim and Codes do not have native support for the
Jellyfish topology. We extended the codes of both simulators to include
Jellyfish. Four path calculation methods (KSP, rKSP, EDKSP, and rEDKSP) are added
for both Booksim and Codes. For Booksim, five routing mechanisms are added:
{\em random}, {\em round-robin}, {\em vanilla UGAL}, {\em KSP-UGAL}, and
{\em KSP-adaptive}. For Codes, two routing mechanisms are added: {\em random} and
{\em KSP-adaptive}.

Booksim parameters are similar to those used in
\cite{Rahman:2019:TUR:3295500.3356208,Kim:2015:HPCA} for simulating HPC systems.
The UGAL routing
techniques are set to have no bias towards MIN or VLB paths. We assume
single-flit packets and a factor 2.0 router speedup. The latency of the channels
are set to 10 cycles. To avoid deadlocks, we increase the virtual channel number
everytime a packet takes a new hop, so the total number of virtual channels
is equal to the diameter of the network.
The buffer size is 32 for each virtual channel.
For each data point, Booksim wormed up for 500 cycles and then collected the
results over a window of 5000 cycles divided into 10 samples each sample of
a window of 500 cycles.  Booksim considers the network to be saturated when
the average packet latency of a sample exceeds 500 cycles. We record the last
injection rate before the network reaches the saturation point as the network
throughput.

Codes has more control parameters than Booksim. We set the router delay, soft
delay, copy per byte and nic delay as 0, and other key parameters to be the same
as those in Booksim to make sure that the evaluation with CODES has no extra
latency than that in Booksim. The link bandwidth is set to 20GBps in the simulation.
The packet size is 1500 Bytes and the buffer size is 64 packets. 

\subsubsection*{Traffic patterns}
On the throughput model, four different traffic patterns are used:
{\em random-permutations} traffic, $Shift(X)$ traffic, $Random(X)$ traffic,
and {\em All to All} traffic. On booksim, three different traffic are used:
{\em random-permutations}, {\em random-shift}, and {\em uniform-random}.
On codes TBA.
With {\em Random(X)} pattern, each processing node sends traffic to X  randomly
picked destinations. With the {\em random-permutations} pattern, each processing
node communicates with at most one other processing node. In a
$Shift(X)$ pattern, a processing node $i$ communicate with proccing
node $(i + X) \ mode \  number\_of\_proceesing\_node$. In the {\em All to All}
pattern, each processing node communicates with all other processing nodes
in the system.  Finally, with {\em uniform-random}, the probability of sending
a packet to each destination is equal.

The communication times of four common HPC communication patterns are evaluated in
Codes: 2D Nearest Neighbor (2DNN), 
2D Nearest Neighbor with diagonal (2DNNdiag),
3D Nearest Neighbor (3DNN), 
3D Nearest Neighbor with diagonal (3DNNdiag).

We collected DUMPI traces~\cite{Dumpi} for the above applications, making sure that
the trace size is same as the network size. For example, for Jellyfish of size 720
routers and 5 compute nodes per router, the network has 3600 compute nodes in total.
So we collected the traces for 3600 ranks and kept the dimension size of 2DNN and
2DNNDiag as $60 \times 60$. Similarly, for 3DNN and 3DNNDiag, we kept the dimension
size as $16\times 15\times 15$.
For all of the applications, each process sends a total of 15MB data, which is divided
among the flows from the process. For example, in 2DNN, each process sends to 4 neighbors.
Thus, each neighbor will receive $\frac{15}{4} = 3.75MB$ of data.
Physical communication patterns are also affected by the process-to-node mapping. In our study,
two mappings for these applications are simulated, linear mapping where processes
are mapped to the nodes in the network sequentially and random mapping where processes are randomly
mapped to the nodes in the network. 

%

\subsection{Properties of the paths computed with different
  path selection schemes}

Before we present our evaluation results, let us first discuss relevant properties of the
paths computed with different path selection schemes. These properties will help us
understand the performance results since 
the paths selected to carry traffic will have a profound impact on the
communication performance.

Table \ref{tab:apl} shows the average 8-shortest path length for RRG(36,24,16),
RRG(720,24,19) and RRG(2880,48,38) with KSP(8), rKSP(8), EDKSP(8), and rEDKSP(8).
Intuitively, the randomization should not statistically increase the average path length
while edge-disjointness will result in longer paths although the exact impact
of the heuristics on the path length depends on the topology. 
The table shows that both randomization and edge-disjointness do not increase the
average path length on RRG(36,24,16) and RRG(2880,48,38), while there is about
$\%4$ average path length increase with EDKSP(8) and rEDKSP(8) on RRG(720,24,16).
Overall, the impact on the path length by the two heuristics is small, which means that
these heuristics will result in better paths without significantly increasing the
path length. 

\begin{table}[htbp]
  \centering
  \begin{tabular}{|p{2.1cm}|p{1.1cm}|p{1.1cm}|p{1.1cm}|p{1.1cm}|}
   \hline
     topology &  \multicolumn{4}{c|}{average K-ptahs length}\\
   
    \cline{2-5}
     & KSP & rKSP & EDKSP & rEDKSP \\
    \hline
     RRG(36,24,16)  & 2.06 & 2.06& 2.06 &2.06\\ 
    \hline
     RRG(720,24,19)  & 3.017 & 3.017 & 3.156 & 3.156\\  
    \hline
     RRG(2880,48,38)  & 2.94  & 2.94 & 2.94 & 2.94 \\ 
    \hline
  \end{tabular}
  \caption{ the average pathes length with K = 8}
  \label{tab:apl}
\end{table}

Tables \ref{tab:count} and \ref{tab:max} give the load-balance property of the $k$-paths computed
with different schemes. Tables \ref{tab:count} show the percentage of
source-desintation switch pairs whose $k$ paths do not share any link. As can be seen in the table,
with KSP and rKSP, the percentage is very low for reasonably large topologies,
while EDKSP and rEDKSP guarantee the paths are link disjoint. With a fixed $k$,
link-disjoint paths offer higher throughput for the source-destination pair than non-link-disjoint
paths for the pair. Hence, the quality of the paths computed by EDKSP and rEDKSP is higher than that
by KSP and rKSP. Table \ref{tab:max} shows the maximum number of paths for a source-destination
pair sharing an edge. With KSP, the value is 6 for the small topology and 7 for the two
larger topologies. For the large topology, this means that there exists at least one pair of switches
whose 7 out of 8 paths share one link. Hence, for this pair, the bandwidth capacity is equivalent to
2 paths instead of 8. The table also shows that randomization does not solve this problem, but
the edge-disjoint heuristic solves the issue. Overall, the data indicate that KSP with
the randomization and edge-disjoint heuristics yield higher quality paths than vanilla KSP. 

\begin{table}[htbp]
  \centering
  \begin{tabular}{|c|r|r|r|r|} 
     \hline
       			& \multicolumn{4}{c|}{ Percentage of pairs whose maximum} \\     
       topology &  \multicolumn{4}{c|} {number of times a link is use is 1 ($k=8$)}\\
     \cline{2-5}
         & KSP(8) & rKSP(8)& EDKSP(8) & rEDKSP(8)  \\
      \hline
     RRG(36,24,16) & 56\% & 59\% & 100\% & 100\% \\
     \hline
     RRG(720,24,19) & 2\% & 3\% & 100\% & 100\% \\
     \hline
      RRG(2880,48,38) & 9\% & 22\% & 100\% & 100\% \\
    \hline
  \end{tabular}
  \caption{Percentage of pairs whose maximum number of times a link is used is 1}
  \label{tab:count}
\end{table}

\begin{table}[htbp]
  \centering
  \begin{tabular}{|c|r|r|r|r|}
     \hline
       			& \multicolumn{4}{c|}{ Maximum number of time the K} \\     
       topology &  \multicolumn{4}{c|} {paths share an edge ($k=8$)}\\
     \cline{2-5}
         & KSP & rKSP& EDKSP & rEDKSP \\
      \hline
     RRG(36,24,16) & 6 & 3 & 1 & 1 \\
     \hline
     RRG(720,24,19) & 7 & 7& 1 & 1 \\
     \hline
      RRG(2880,48,38) & 7 & 6 & 1 & 1 \\
    \hline
  \end{tabular}
  \caption{Maximum number of times an edge has been shared by a single SD K-paths ($k=8$).}
  \label{tab:max}
\end{table}


\subsection{Throughput modeling Results}

\begin{figure}[htbp]
  \centering
        \includegraphics[width=2.8in]{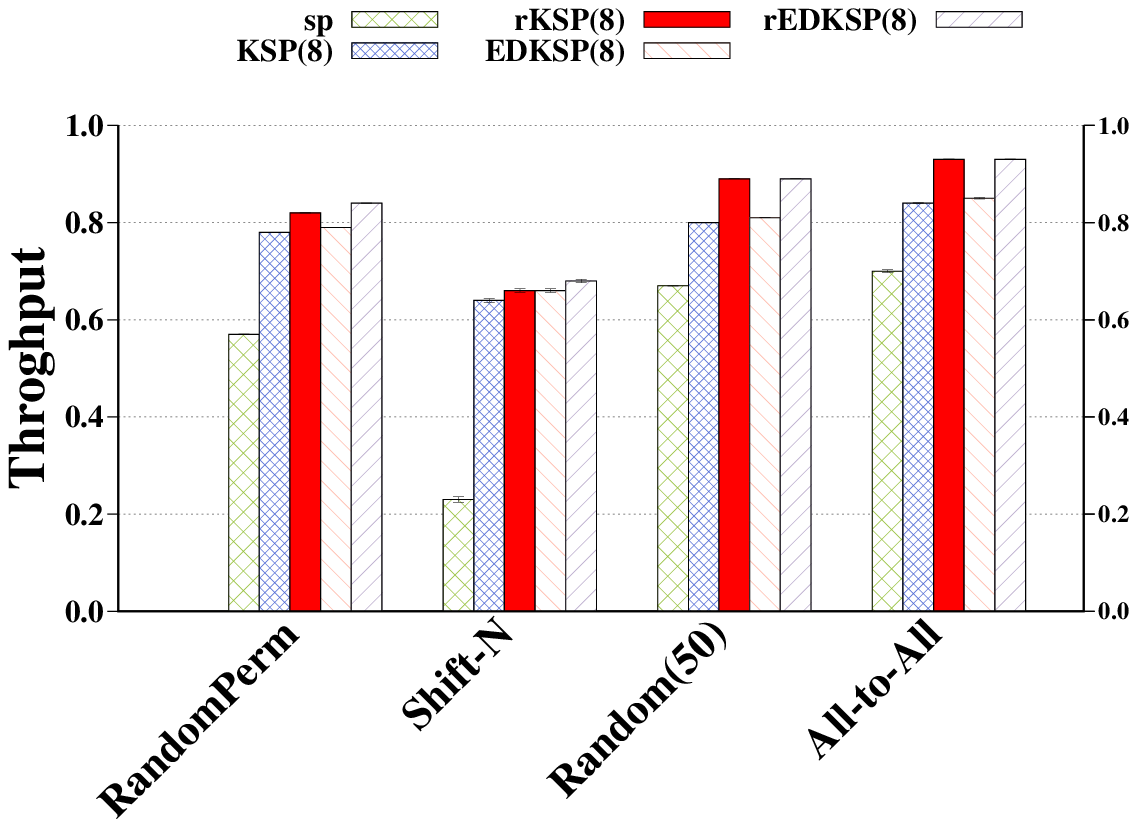}
        \caption{The average model throughput on
          RRG(36, 16, 8)}
        \label{fig:m36}
\end{figure}

\begin{figure}[htbp]
  \centering
        \includegraphics[width=2.8in]{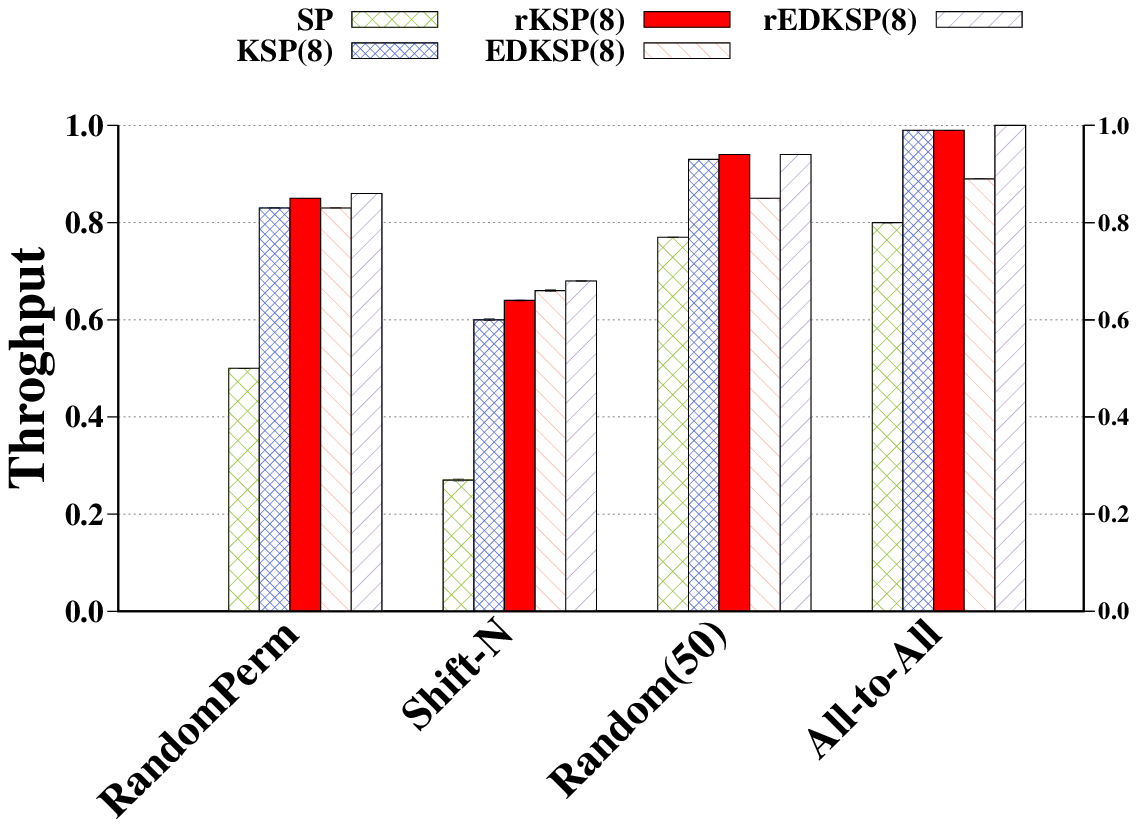}
        \caption{The average model throughput on
          RRG(720, 24, 19)}
        \label{fig:m720}
\end{figure}

\begin{figure}[htbp]
  \centering
        \includegraphics[width=2.8in]{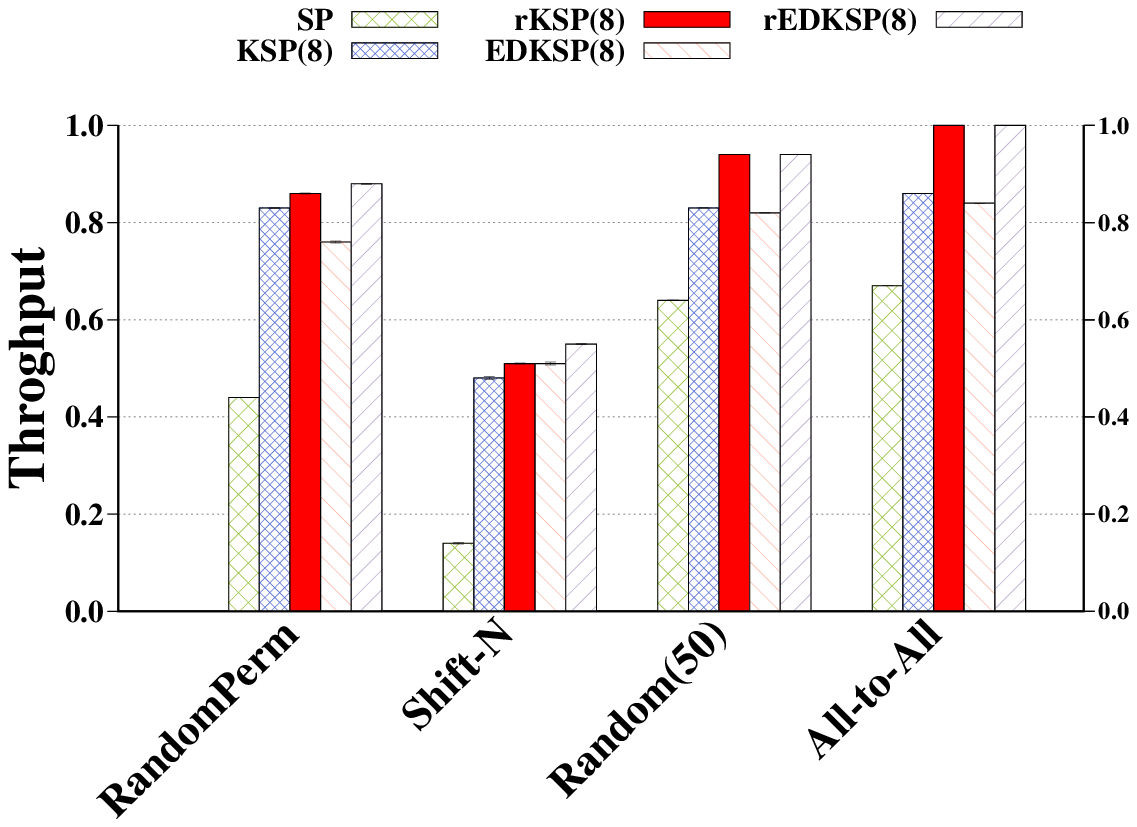}
        \caption{The average model throughput on
          RRG(2880, 48, 38)}
        \label{fig:m2880}
\end{figure}

Figures \ref{fig:m36}, \ref{fig:m720}, and \ref{fig:m2880} shows the
average modeling throughput per flow for RRG(36,24,16), RRG(720,24,19) and,
RRG(2880,48,38) respectivly, on four different traffic patterns. 

For each topology, we created 10 random samples. For shift traffic,
random-permutations traffic, and random(x=50) traffic, 50 different random
instances have been used per topology sample.  The average of the random samples
for four routing schemes KSP(k=8), rKSP(k=8), EDKSP(k=8), and rEDKSP(k=8) are reported.
The throughput value in the figure is the per node normalized throughput. A value of
1 means the flow can communicate at the full link speed; a value of 0.8 means that the
flow can communicate at 80\% of the link speed. 
There are several interesting observations in these figures.
First, randomization has significant impacts on both KSP and EDKSP, and noticeably
improve the modeled throughput.
For example, for RRG(36,24,16) figure \ref{fig:m36} for Random(50) traffic,
the average throughput for KSP(8) is 0.80 while for
rKSP(8) is .89, 10\% higher. For RRG(2880,24,16)
in Figure \ref{fig:m2880}, the average model throughput for random-permutation
traffic with EDKSP(8) is 0.76 while that with rEDKSP is 0.88, 14\% higher.
The results indicate that systematic biases in path selection
can cause performance issues on Jellyfish and
randomization is an effective method to alleviate the problem.

Second, $rEDKSP$ consistently achieves the highest performance,
out-performing all other path selection schemes. For example, for random permutation
on RRG(36,24,16) in Figure~\ref{fig:m36}, the average throughput
is 0.82 for rKSP(8) and 0.86 for rEDKSP(8), 5\% higher. For shift traffic on RRG(2880,24,16)
in Figure \ref{fig:m2880}, the average throughput is 0.51 for rKSP(8) and 0.55 for
rEDKSP(8), 7\% higher. In our more detailed simulation,
rEDKSP out-performs rKSP even more, which indicates that using edge-disjoint path
is effective in achieving
load balancing and rEDKSP is the best performing path calculation method for multi-path
routing on Jellyfish.

Finally, multi-path routing schemes consistently out-perform single-path routing to a large
degree, which confirms earlier results \cite{singla2012jellyfish,yuan2013new}.
  
%
%

\subsection{Results from Booksim}

As mentioned earlier, with the cycle-level simulation, Booksim results are more accurate than
the model. Using Booksim, we investigated the performance of KSP, rKSP, EDKSP, and rEDKSP
with different routing mechanisms for two common traffic patterns: random permutation and random
shift.

\begin{figure}[htbp]
  \centering
        \includegraphics[width=2.8in]{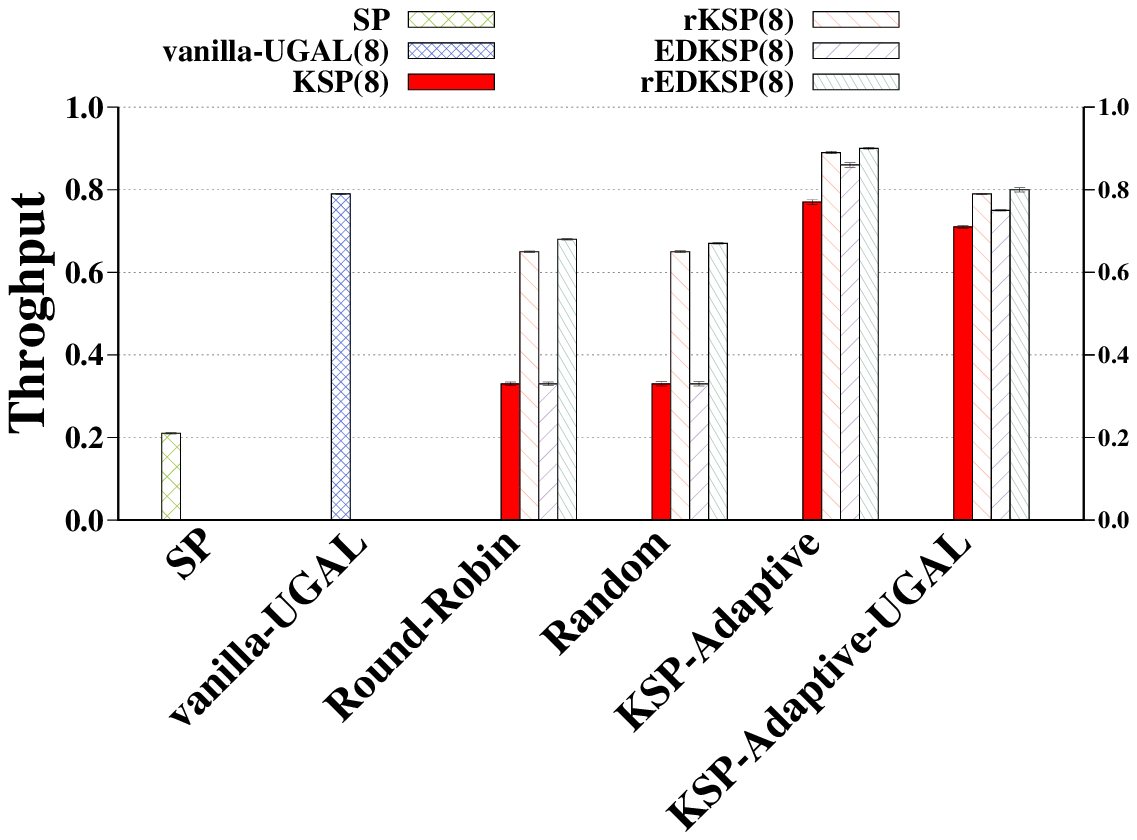}
        \caption{The average throughput of random permutations on
          RRG(36, 24, 16)}
        \label{fig:B36_perm}
\end{figure}

\begin{figure}[htbp]
  \centering
        \includegraphics[width=2.8in]{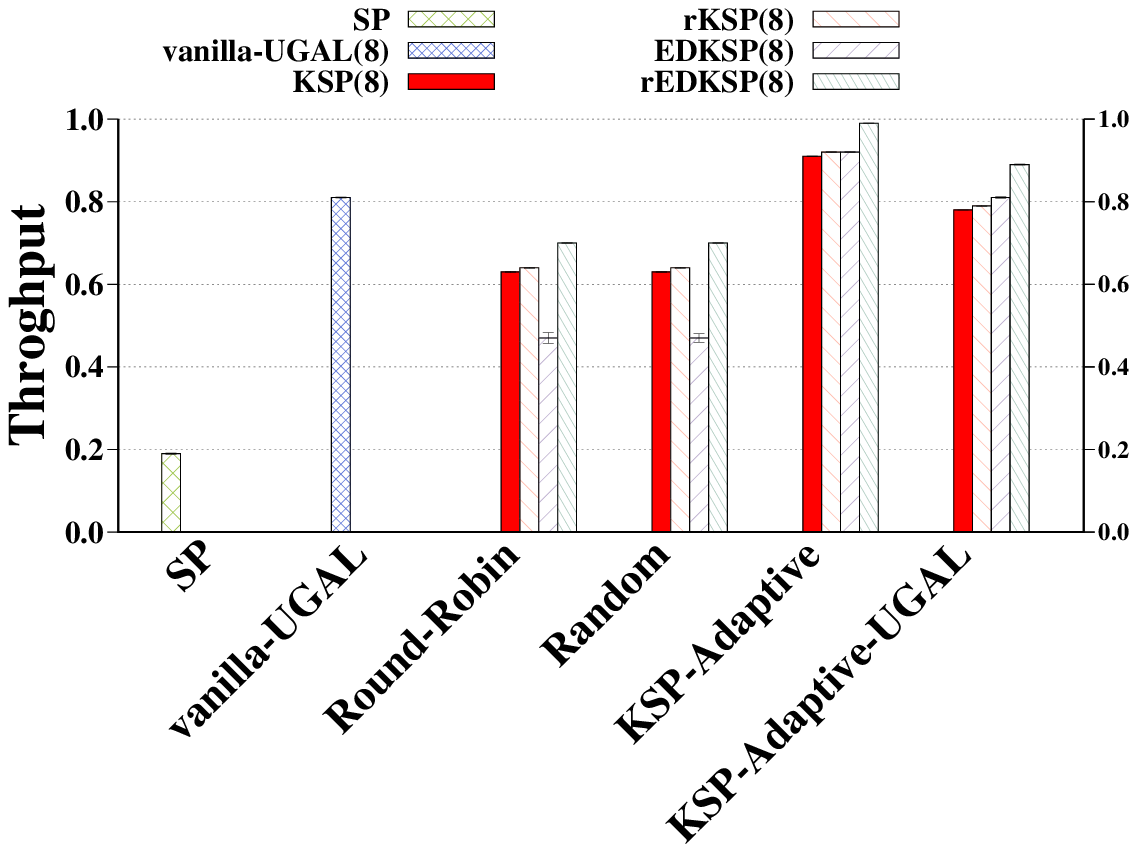}
        \caption{The average throughput of random permutations on
          RRG(720, 24, 19)}
        \label{fig:B720_perm}
\end{figure}

Figure~\ref{fig:B36_perm} and \ref{fig:B720_perm} shows the average saturation throughput for
the random permutation pattern with different path calculation and different routing mechanisms.
The results are the average of ten random patterns. 
We observe the following. First, across different routing mechanisms,
rEDKSP consistently achieves the highest performance among the path selection schemes.
The throughput is very signficantly better than that for KSP. For example, using KSP-adaptive,
on RRG(720, 24, 19), rEDKSP achieves a throughput of 0.98, 11\% higher than the
throughput of 0.88 for KSP. This demonstrates the effectiveness of
randomization and edge-disjointness. 
Second, the routing mechanim has a significant impact on the performance, traffic adaptive
routing (KSP-adaptive and KSP-UGAL) performs better than traffic oblivious routing.
Between the adaptive routing scheme, KSP-adaptive is significantly better than KSP-UGAL for all
path selection schemes. For example, with rEDKSP, KSP-adaptive achieves a throughput of 0.98,
10\% higher than the throughput of 0.89 for KSP-UGAL. Moreover, KSP-adaptive and KSP-UGAL both
have higher performance than the vanilla UGAL although vanilla UGAL is more flexible in selecting
non-minimal paths. Restricting the paths to be ``short'' paths results in higher performance
on Jellyfish. 

\begin{figure}[htbp]
  \centering
        \includegraphics[width=2.8in]{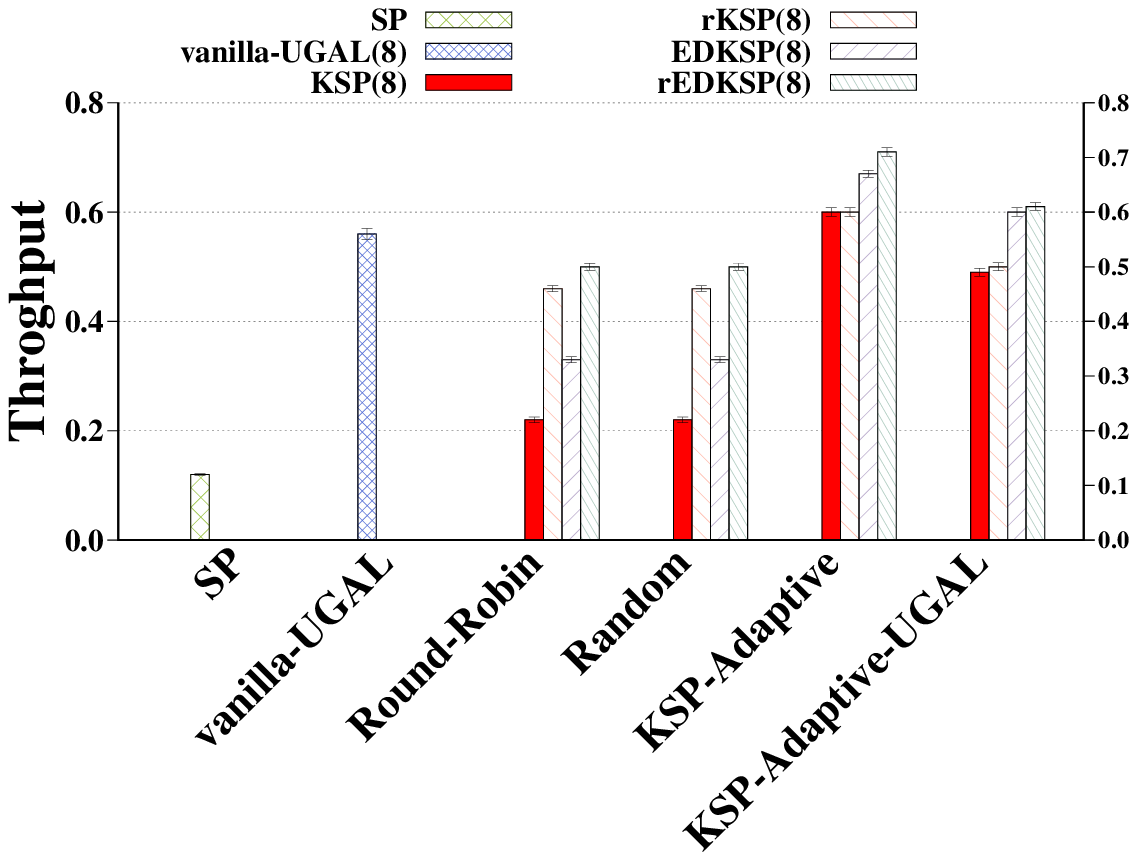}
        \caption{The average throughput of random shift on
          RRG(36, 24, 16)}
        \label{fig:B36_shift}
\end{figure}

\begin{figure}[htbp]
  \centering
        \includegraphics[width=2.8in]{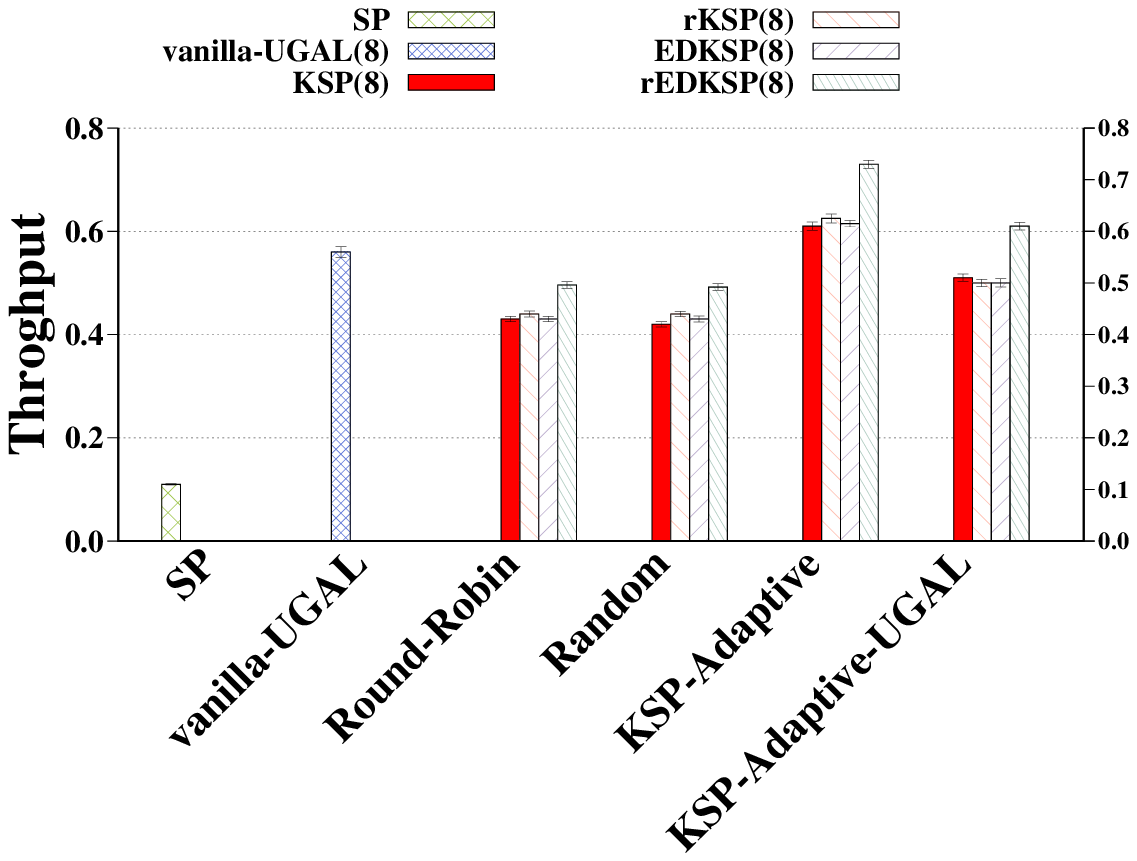}
        \caption{The average throughput of random shift on
          RRG(720, 24, 19)}
        \label{fig:B720_shift}
\end{figure}

Figure~\ref{fig:B36_shift}, and \ref{fig:B720_shift} shows the average saturation
throughput for random shift patterns with different
path calculation and different routing schemes. The results are similar to that
for permutation except that the difference is larger. This is because an average shift
traffic pattern is more demanding than an average permutation. KSP-adaptive is the best
performing routing mechanism. With rEDKSP(8), KSP-adaptive achieves a throughput of 0.72,
20\% higher than the 0.6 throughput for KSP-UGAL. With KSP-adaptive, rEDKSP(8) achieves
a throughput of 0.72, 20\% higher than the 0.6 throughput with KSP(8). 

\begin{figure}[htbp]
  \centering
        \includegraphics[width=2.8in]{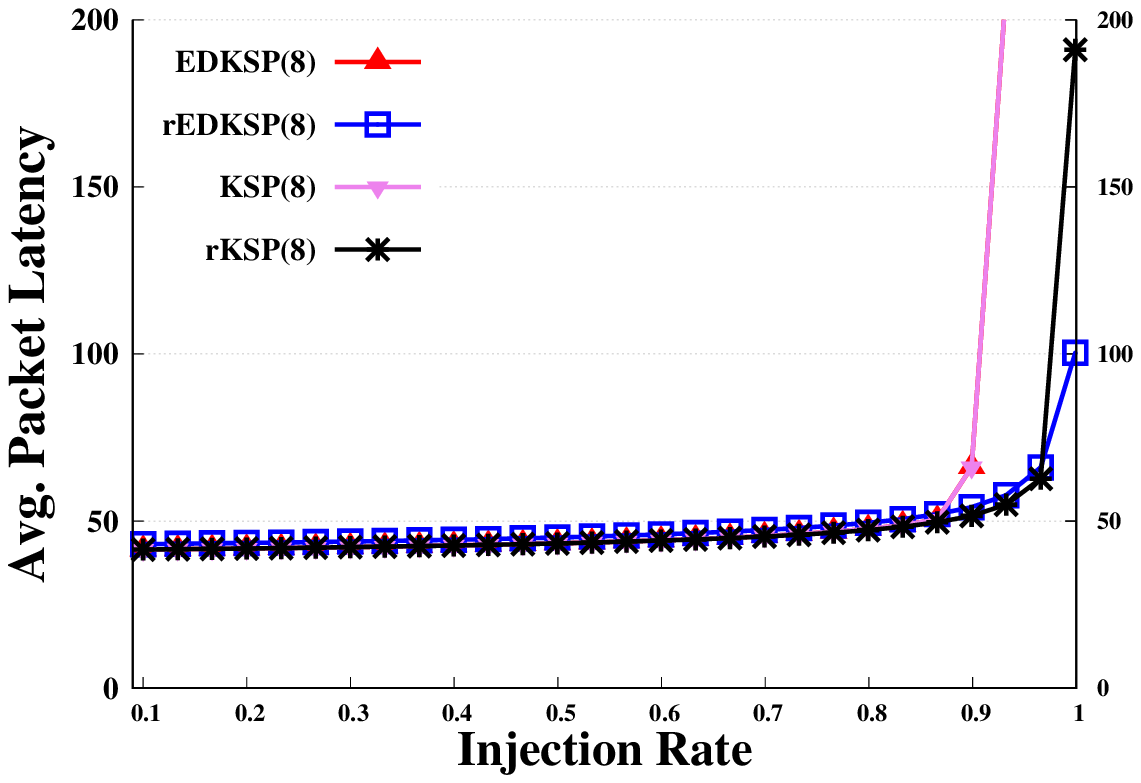}
        \caption{The average packet latency of random uniform on
          RRG(720, 19, 5) on random routing scheme}
        \label{fig:B720U_lat}
\end{figure}

\begin{figure}[htbp]
  \centering
        \includegraphics[width=2.8in]{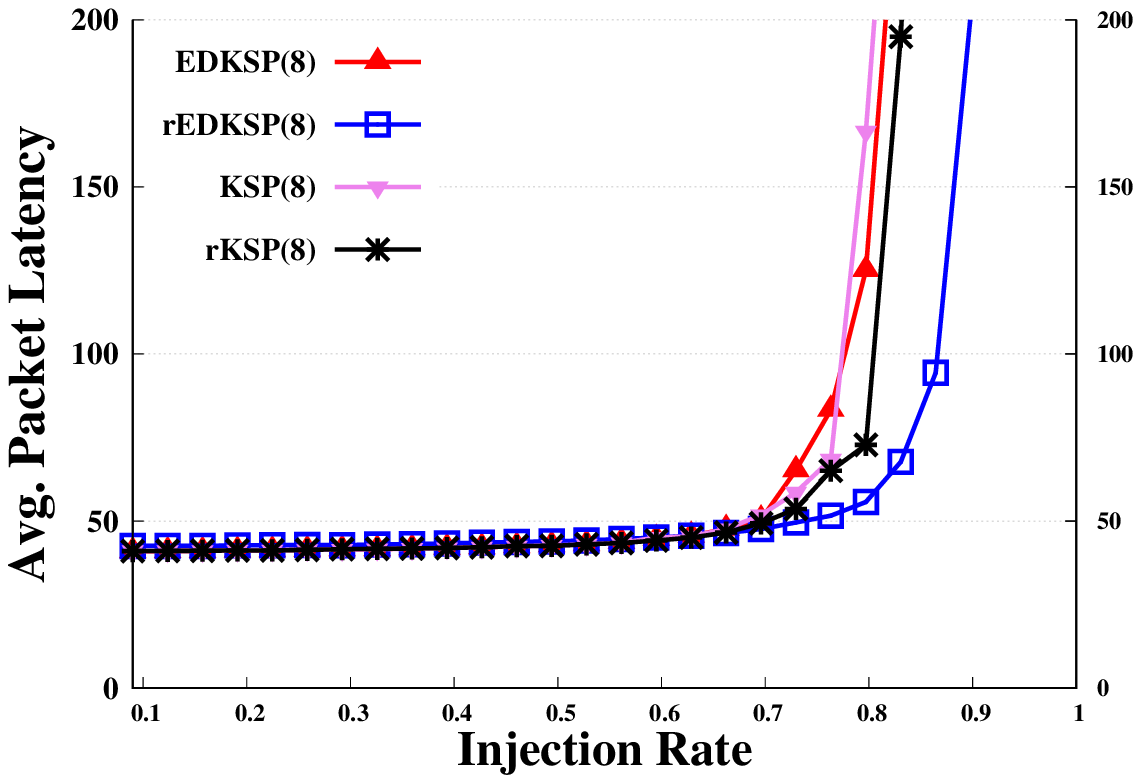}
        \caption{The average packet latency of random permutation on
          RRG(720, 24, 19) on adaptive routing scheme}
        \label{fig:B720P_lat}
\end{figure}

\begin{figure}[htbp]
  \centering
        \includegraphics[width=2.8in]{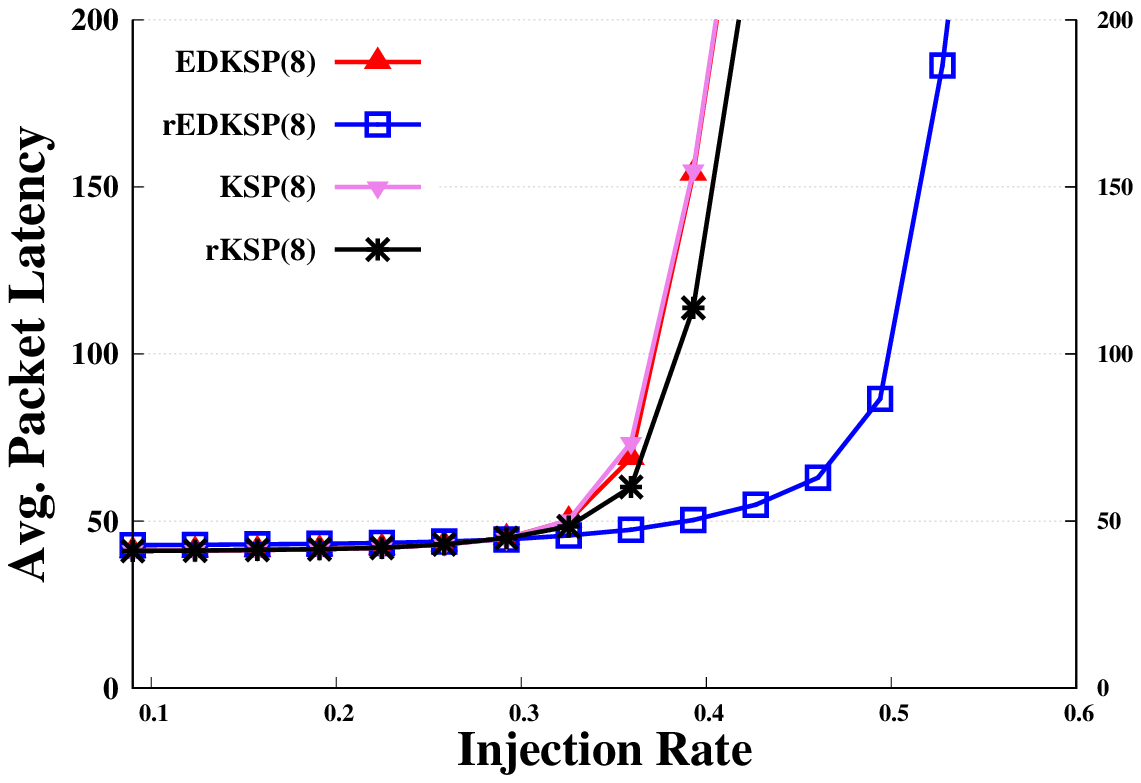}
        \caption{The average packet latency of shift(X) on
          RRG(720, 24, 19) on adaptive routing scheme}
        \label{fig:B720S_lat}
\end{figure}

Figure \ref{fig:B720U_lat}, \ref{fig:B720P_lat} and \ref{fig:B720S_lat}
show the average packet latency as the offered load increases for RRG(720,19,5) on three
traffic conditions, random-uniform traffic, a random permutation traffic, and a random shift traffic,
respectively. The routing mechanism is KSP-adaptive.
The results are consistent with all three traffic patterns. At low loads, all path
selection schemes have similar latency. At high loads, rEDKSP achieves higher throughput
and lower latency near saturation. 
This is because randomization and edge-disjointedness
heuristics provide a better load-balance compared to KSP. The results demonstrate the effectiveness
of rEDKSP, especially in comparison to the vanilla KSP. 


\subsection{Results from Codes}

The communication times for Stencil communication patterns, which are very common in
HPC application, are studied using Codes. Such results give a good indication about how the
network performs in a more realistic setting. 
Tables~\ref{tab:linear_map} shows the average
communication times for each of the four applications with linear process-to-node mapping; 
The experiments are on $RRG(720, 24, 19)$, the link bandwidth is set to 20GBps.
For each application, each process sends a total of 15MB data. The second column is the
communication time with rEDKSP(8).
The third column is the communication with KSP(8).
The fourth column is the improvement percentage of rEDKSP(8) over KSP(8).
The fifth column is the communication with rKSP(8).
The sixth column is the improvement percentage of rEDKSP(8) over rKSP(8).
With linear mapping, rEDKSP out-performs KSP and rKSP for all four Stencil patterns. On average,
rEDKSP(8) improves over KSP(8) by 7.6\% and over rKSP(8) by 4.5\%. This is consistent with our
results from the model and Booksim.

\begin{table}[htbp]
  \centering
  \begin{tabular}{|r|r|r|r|r|r|}
	\hline
	Applications & rEDKSP(8) & \multicolumn{2}{c|}{KSP(8)} & \multicolumn{2}{c|}{rKSP(8)} \\
        \cline{3-6}
                     & time      & time & imp. & time & imp.\\ 
	\hline
	2DNN &  0.83 & 0.91 & 9.6\% & 0.88 & 6.0\%\\
	\hline
	2DNNdiag &  1.07 & 1.20 & 12.1\% & 1.15 & 7.5\%\\
	\hline
	3DNN &  0.90 & 0.95 & 5.6\% & 0.93 & 3.3\%\\
	\hline
	3DNNdiag  & 1.01 & 1.04 & 3.0\% & 1.02 & 1.0\% \\
        \hline
        Average   &      &      & 7.6\% & & 4.5\%\\
	\hline
  \end{tabular}
  \caption{Communication time in milliseconds for different routing schemes with linear mapping of processes to nodes on RRG(720, 24, 19)}
  \label{tab:linear_map}
\end{table}

Table~\ref{tab:random_map} shows the results with random process-to-node mapping.
The trend is similar to that with the linear mapping. On average rEDKSP(8) out-performs
KSP(8) by 9.0\%, and rKSP(8) by 0.8\%. For 2DNNdiag, rEDKSP(8) performs slightly
worse than rKSP(8). For the random mapping, rEDKSP performs very similar to rKSP. We attribute this
to the random traffic pattern resulted from the random mapping. Overall, rEDKSP is still slightly
better rKSP in this condition. In summary, for the Stencil patterns,
rEDKSP consistently achieves higher performance than other path selection methods. 

\begin{table}[htbp]
  \centering
  \begin{tabular}{|r|r|r|r|r|r|r|}
	\hline
	Applications & rEDKSP(8) & \multicolumn{2}{c|}{KSP(8)} & \multicolumn{2}{c|}{rKSP(8)} \\
        \cline{3-6}
                             & time      & time & imp. & time & imp.\\ 
	\hline
	2DNN  & 0.92 & 0.99 & 7.6\% & 0.94 & 2.2\% \\
	\hline
	2DNNdiag  & 0.86 & 0.92 & 7.0\% & 0.84 & -1.5\%\\
	\hline
	3DNN  & 0.88 & 0.95 & 8.0\%& 0.88 & 0.0\%\\
	\hline
	3DNNdiag  & 0.76 & 0.86 & 13.2\% & 0.78 & 2.6\%\\
	\hline
        Average   &     &      & 9.0\%   &      & 0.8\%\\
        \hline
  \end{tabular}
  \caption{Communication time in milliseconds for different routing schemes in random mapping of processes to nodes one RRG(720, 24, 19)}
  \label{tab:random_map}
\end{table}

\section{Related work}
\label{related}

Singla et al. show the Jellyfish can be more cost-effective than fat-tree
\cite{singla2012jellyfish}. Follow-up studies conclude that Jellyfish
is effective \cite{jyothi2016measuring}, and is more scalable than
fat-tree for HPC systems \cite{yuan2013new}. It is known that single path routing
and equal-cost multiple path rotuing  do not work well on Jellyfish,
and KSP has been suggested for Jellyfish \cite{singla2012jellyfish,yuan2013new}.
KSP allows other heuristics, such as randomization and link-disjointness, to
be incorproated. Such heuristics, however, have not been examined on the Jellyfish
topology, which has unique topological properties. As a result, it is not
clear whether they are effective on the Jellyfish topology.
Our study indicates that, for the Jellyfish topology, the randomization
and link-disjoint heuristics yield better paths, and significantly
improves the routing performance. Routing mechanisms for multi-path routing on Jellyfish
have not been systematically investigated. An ad hoc study was carried out where
Jellyfish is compared with other topologies \cite{mollah2018comparative}. 
In this work, we study different forms of applying the Universal Globally Adaptive Load-balanced
routing (UGAL) \cite{singh2005load} to multi-path routing on Jellyfish. Like the Dragonfly
topology where UGAL has many forms
\cite{Kim:2008:ISCA:Dragonfly,Jiang:2009:ISCA:Indirect_adaptive,Faizian:2018:TMSCS,Mollah:2019:TOPC,Rahman:2019:TUR:3295500.3356208}, multi-path routing on Jellyfish also allows for UGAL variations.
We evaluate their performance, compare the performance with that of single path
routing and traffic oblivious multi-path routing, and identify the most effective routing
mechanism for Jellyfish.




\section{Conclusion}
\label{conc}

We study two components of multi-path routing on Jellyfish, path selection and routing
mechanism. We show that the current KSP routing for Jellyfish suffers from
the load imbalance problem. We introduce two heuristics, randomization and edge-disjointness,
to overcome this issue, and demonstrate that these two heuristics yield significantly
better performance than the vanilla KSP scheme. We investigate various routing mechanisms
for multi-path routing on Jellyfish. Our results indicate that using paths computed
by KSP with the randomization and edge-disjointness heuristics and the KSP-adaptive
scheme significantly improve the communication performance on Jellyfish in comparison
to existing schemes. 


\bibliographystyle{abbrv}
\bibliography{201}

\begin{thebibliography}{10}

\bibitem{Cope:2011:Codes}
J.~Cope, N.~Liu, S.~Lang, P.~Carns, C.~Carothers, and R.~Ross.
\newblock Codes: Enabling co-design of multilayer exascale storage
  architectures.
\newblock In {\em the Workshop on Emerging Supercomputing Technologies}, 2012.

\bibitem{Faizian:2018:TMSCS}
P.~Faizian, J.~F. Alfaro, M.~S. Rahman, M.~A. Mollah, X.~Yuan, S.~Pakin, and
  M.~Lang.
\newblock {TPR:} traffic pattern-based adaptive routing for dragonfly networks.
\newblock {\em {IEEE} Trans. Multi-Scale Computing Systems}, 4(4):931--943,
  2018.

\bibitem{guo2003link}
Y.~Guo, F.~Kuipers, and P.~Van~Mieghem.
\newblock Link-disjoint paths for reliable qos routing.
\newblock {\em International Journal of Communication Systems}, 16(9):779--798,
  2003.

\bibitem{Han06}
H.~{Han}, S.~{Shakkottai}, C.~V. {Hollot}, R.~{Srikant}, and D.~{Towsley}.
\newblock Multi-path tcp: A joint congestion control and routing scheme to
  exploit path diversity in the internet.
\newblock {\em IEEE/ACM Transactions on Networking}, 14(6):1260--1271, 2006.

\bibitem{hoffman1959method}
W.~Hoffman and R.~Pavley.
\newblock A method for the solution of the n th best path problem.
\newblock {\em Journal of the ACM (JACM)}, 6(4):506--514, 1959.

\bibitem{Dumpi}
C.~L. Janssen, H.~Adalsteinsson, S.~Cranford, J.~P. Kenny, A.~Pinar, D.~A.
  Evensky, and J.~Mayo.
\newblock A simulator for large-scale parallel computer architectures.
\newblock {\em International Journal of Distributed Systems and Technologies
  (IJDST)}, 1(2):57--73, 2010.

\bibitem{Jiang:2013:booksim}
N.~{Jiang}, D.~U. {Becker}, G.~{Michelogiannakis}, J.~{Balfour}, B.~{Towles},
  D.~E. {Shaw}, J.~{Kim}, and W.~J. {Dally}.
\newblock A detailed and flexible cycle-accurate network-on-chip simulator.
\newblock In {\em 2013 IEEE International Symposium on Performance Analysis of
  Systems and Software (ISPASS)}, pages 86--96, April 2013.

\bibitem{Jiang:2009:ISCA:Indirect_adaptive}
N.~Jiang, J.~Kim, and W.~J. Dally.
\newblock Indirect adaptive routing on large scale interconnection networks.
\newblock In {\em Proceedings of the 36th Annual International Symposium on
  Computer Architecture}, ISCA '09, pages 220--231, New York, NY, USA, 2009.
  ACM.

\bibitem{jyothi2016measuring}
S.~A. Jyothi, A.~Singla, P.~B. Godfrey, and A.~Kolla.
\newblock Measuring and understanding throughput of network topologies.
\newblock In {\em SC'16: Proceedings of the International Conference for High
  Performance Computing, Networking, Storage and Analysis}, pages 761--772.
  IEEE, 2016.

\bibitem{Kim:2008:ISCA:Dragonfly}
J.~Kim, W.~J. Dally, S.~Scott, and D.~Abts.
\newblock Technology-driven, highly-scalable dragonfly topology.
\newblock In {\em Proceedings of the 35th Annual International Symposium on
  Computer Architecture}, ISCA '08, pages 77--88, Washington, DC, USA, 2008.
  IEEE Computer Society.

\bibitem{martins2003new}
E.~Q. Martins and M.~M. Pascoal.
\newblock A new implementation of yen’s ranking loopless paths algorithm.
\newblock {\em Quarterly Journal of the Belgian, French and Italian Operations
  Research Societies}, 1(2):121--133, 2003.

\bibitem{mollah2018comparative}
M.~A. Mollah, P.~Faizian, M.~S. Rahman, X.~Yuan, S.~Pakin, and M.~Lang.
\newblock A comparative study of topology design approaches for hpc
  interconnects.
\newblock In {\em 2018 18th IEEE/ACM International Symposium on Cluster, Cloud
  and Grid Computing (CCGRID)}, pages 392--401. IEEE, 2018.

\bibitem{Mollah:2019:TOPC}
M.~A. Mollah, W.~Wang, P.~Faizian, M.~S. Rahman, X.~Yuan, S.~Pakin, and
  M.~Lang.
\newblock Modeling universal globally adaptive load-balanced routing.
\newblock {\em ACM Trans. Parallel Comput.}, 6(2), Aug. 2019.

\bibitem{Rahman:2019:TUR:3295500.3356208}
M.~S. Rahman, S.~Bhowmik, Y.~Ryasnianskiy, X.~Yuan, and M.~Lang.
\newblock Topology-custom ugal routing on dragonfly.
\newblock In {\em Proceedings of the International Conference for High
  Performance Computing, Networking, Storage and Analysis}, SC '19, pages
  17:1--17:15, New York, NY, USA, 2019. ACM.

\bibitem{singh2005load}
A.~Singh.
\newblock {\em Load-balanced routing in interconnection networks}.
\newblock PhD thesis, Stanford University, 2005.

\bibitem{singla2012jellyfish}
A.~Singla, C.-Y. Hong, L.~Popa, and P.~B. Godfrey.
\newblock Jellyfish: Networking data centers randomly.
\newblock In {\em Presented as part of the 9th $\{$USENIX$\}$ Symposium on
  Networked Systems Design and Implementation ($\{$NSDI$\}$ 12)}, pages
  225--238, 2012.

\bibitem{Kim:2015:HPCA}
J.~Won, G.~Kim, J.~Kim, T.~Jiang, M.~Parker, and S.~Scott.
\newblock Overcoming far-end congestion in large-scale networks.
\newblock In {\em High Performance Computer Architecture (HPCA), 2015 IEEE 21st
  International Symposium on}, pages 415--427, Feb 2015.

\bibitem{yen1971finding}
J.~Y. Yen.
\newblock Finding the k shortest loopless paths in a network.
\newblock {\em management Science}, 17(11):712--716, 1971.

\bibitem{yuan2013new}
X.~Yuan, S.~Mahapatra, W.~Nienaber, S.~Pakin, and M.~Lang.
\newblock A new routing scheme for jellyfish and its performance with hpc
  workloads.
\newblock In {\em Proceedings of the International Conference on High
  Performance Computing, Networking, Storage and Analysis}, pages 1--11, 2013.

\end{thebibliography}

\end{document}